\documentclass[jcp,aip,10pt,showpacs,reprint,floatfix]{revtex4-1}
\usepackage{amsmath,amssymb,amsfonts,graphicx}

\newcommand{\IP}{\alpha_{\rm I}}
\newcommand{\NF}{N_{\rm f}}
\newcommand{\NX}{N_\times}
\newcommand{\KF}{\kappa_{\rm f}}
\newcommand{\KS}{k_{\rm s}}
\newcommand{\KX}{k_\times}
\newcommand{\KR}{\KX/\KS}

\newcommand{\D}{\displaystyle}

\newcommand{\eq}[1]{Eq.~(\ref{#1})}
\newcommand{\fig}[1]{Fig.~\ref{#1}}

\newcommand{\sect}[1]{Section~\ref{#1}}
\newcommand{\avg}[1]{\langle #1 \rangle}
\newcommand{\olcite}[1]{Ref.~\onlinecite{#1}}
\newcommand{\ahum}[1]{``#1''}

\begin{document}

\title{Crosslinked biopolymer bundles: crosslink reversibility leads to 
cooperative binding/unbinding phenomena}

\author{Richard L. C. Vink}
\email{rlcvink@gmail.com}
\affiliation{Institute of Theoretical Physics, Georg-August-Universit\"at
G\"ottingen, Friedrich-Hund-Platz~1, 37077 G\"ottingen, Germany}

\author{Claus Heussinger$^{1,}$} 
\email{heussinger@theorie.physik.uni-goettingen.de}
\affiliation{Institute of Theoretical Physics, Georg-August-Universit\"at
G\"ottingen, Friedrich-Hund-Platz~1, 37077 G\"ottingen, Germany}
\affiliation{Max Planck Institute for Dynamics and Self-Organization, 
Bunsenstra{\ss}e 10, 37073 G\"ottingen, Germany}

\begin{abstract} We consider a biopolymer bundle consisting of filaments that 
are crosslinked together. The crosslinks are reversible: they can dynamically 
bind and unbind adjacent filament pairs as controlled by a binding enthalpy. The 
bundle is subjected to a bending deformation and the corresponding distribution 
of crosslinks is measured. For a bundle consisting of two filaments, upon 
increasing the bending amplitude, a first-order transition is observed. The 
transition is from a state where the filaments are tightly coupled by many bound 
crosslinks, to a state of nearly independent filaments with only a few bound 
crosslinks. For a bundle consisting of more than two filaments, a series of 
first-order transitions is observed. The transitions are connected with the 
formation of an interface between regions of low and high crosslink densities. 
Combining umbrella sampling Monte Carlo simulations with analytical 
calculations, we present a detailed picture of how the competition between 
crosslink shearing and filament stretching drives the transitions. We also find 
that, when the crosslinks become soft, collective behavior is not observed: the 
crosslinks then unbind one after the other leading to a smooth decrease of the 
average crosslink density.\end{abstract}



\maketitle

\section{Introduction}

The cytoskeleton is a complex meshwork made of long elastic filaments coupled 
together with the help of numerous, rather compact crosslinking 
proteins~\cite{alb94}. An important aspect of cytoskeletal assemblies is their 
dynamic nature, which allows them to react to external stimuli and adapt their 
internal structure and mechanical properties according to the needs of the cell. 
The reversible nature of crosslink binding is an important mechanism that 
underlies these dynamical processes. For example, living cells show complex 
rheological properties that range from fluidization to reinforcement under 
stress\cite{trepat2007Nature, Fernandez2006BPJFibroblast}, and reversible bonds 
between cytoskeletal filaments have been proposed as key mechanisms in mediating 
between these contradicting behaviors \cite{KollmannsbergerReview2011, 
wolffNJP2010}. Similar effects are believed to be important for the rheological 
properties of reconstituted F-actin networks\cite{bausch06, 
LielegSoftMatter2010, KaszaCOCB2007}, in particular at low frequencies that 
correspond to the lifetime of the crosslink-mediated 
bond\cite{lielegPRL2008Transient, broederszPRL2010Linker}.

Another important class of cytoskeletal assemblies are filament bundles. 
Crosslinked F-actin bundles form primary structural components of a broad range 
of cytoskeletal structures including stereocilia, filopodia, microvilli or the 
sperm acrosome. Type and properties of the crosslinking protein allow the cell 
to tailor the dimensions and mechanical properties of the bundles to suit 
specific biological functions. In particular, the mechanical properties of these 
bundles play key roles in cellular functions ranging from 
locomotion\cite{mogilnerBPJ2005, atilganBPJ2006, vignejvic2006JCellBiol} to 
mechanotransduction\cite{hudspeth1977PNAS}, and 
fertilization\cite{schmid2004Nature}.

In-vitro experiments and modeling have emphasized the role of the crosslink 
stiffness in mediating bundle mechanical properties\cite{claessens06NatMat, 
Bathe20082955, PhysRevLett.103.238102}. It is less clear, however, in how far 
the reversibility of the crosslinking bond may affect bundle mechanical or 
dynamical properties. On the one hand, one expects reversible bonds to influence 
the conformational properties of the filaments. Examples for this dependency are 
the formation of kinks in the bundle contour\cite{cohenPNAS2003, 
DogicPRL2006Kinks}, or an unbundling transition as the binding affinity of the 
crosslinks is reduced\cite{benetatosPRE2003, kierfeld05PRL}. Conversely, bundle 
conformational changes or the application of destabilizing 
forces\cite{kierfeldPRL2006Unzipping} directly influence the binding state of 
the crosslinks.

The aim of this article is to deepen our understanding of this complex interplay 
between reversible crosslink binding and bundle mechanical and dynamical 
properties. We consider the nonlinear response of a reversibly crosslinked 
filament bundle to an imposed external force or deformation. In particular, we 
want to determine how the external driving is reflected in the internal degrees 
of freedom of the bundle, notably the binding state of the crosslinks. Combining 
simulations and theory we will show that, depending on the mechanical stiffness 
of the crosslinking agent, the fraction of bound crosslinks can display a sudden 
and discontinuous drop. This indicates a {\it cooperative unbinding process} 
that involves the crossing of a free energy barrier. Choosing the proper 
crosslinking protein, therefore, not only allows one to change the composite 
elastic properties of the bundle, but also the relevant time-scales: the latter 
can be tuned from the single crosslink binding rate to the (much longer) escape 
time over the free energy barrier (a short communication of these results was 
recently presented by one of us in \olcite{PhysRevE.83.050902}). We emphasize 
that related effects may be important in a variety of different biological 
contexts also. For example, cellular adhesion and locomotion are dependent on 
the formation of transient cell-to-matrix 
bonds\cite{julianoCellAdhesionReview2002, Critchley2000FocalAdhesion}. Simple 
theoretical models highlight the complex interplay between specific adhesion, 
mediated by the binding agent, unspecific interaction with the substrate, and 
cell-membrane elasticity\cite{BruinsmaSackmannPRE2000, 
weiklSoftMatterReview2009, SmithSeifertSoftMatter2007, weil.faragoEPJE2010}.

The outline of the paper is as follows. In \sect{sec:model}, we define and 
motivate our bundle model. Next, in \sect{sec:sim}, we present our numerical 
results obtained in Monte Carlo simulations. In \sect{sec:the}, the development 
of a theoretical model is described.  The combined efforts of simulation and 
theory allow us to obtain a detailed understanding of the underlying physical 
mechanisms that affect the crosslink binding state. We end in \sect{sec:final} 
with a discussion of the implications and the experimental relevance of our 
results.

\section{Model}
\label{sec:model}
\begin{figure}
\begin{center}
\includegraphics[width=\columnwidth]{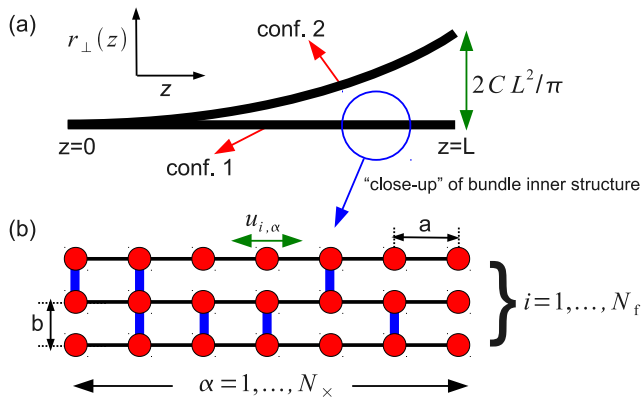}
\caption{\label{fig:illustration1} (a) An end-grafted bundle of length $L$ is 
brought from an initial unbend state (configuration~1) to a bend state 
(configuration~2). The bundle contour $r_\perp(z)$ is given by \eq{eq:shape}, 
where $C$ sets the bending amplitude. (b) \ahum{Close-up} view of the bundle 
inner structure. In our model, the bundle consists of $\NF$ filaments, each 
filament being a chain of $\NX$ beads connected by harmonic springs (horizontal 
bonds). The beads can slide along the bundle contour, as indicated by the 
displacement $u_{i,\alpha}$. The filaments are joined to each other by 
crosslinks (vertical bonds) which may dynamically bind and unbind.}
\end{center}
\end{figure}

We consider a bundle in a two-dimensional plane (\fig{fig:illustration1}(a)). 
The bundle has length $L$ and, in the initial state, is oriented along the 
$z$-axis of a fixed laboratory frame (configuration~1). We now envision an 
experiment whereby the bundle is brought from the initial state to a \ahum{bent} 
state (configuration~2). In the bent state, the contour (shape) of the bundle is 
described by a transverse displacement $r_\perp(z)$, the functional form of 
which depends on the boundary conditions and the specific way of loading 
(e.g.~bending or buckling). While the details of the loading are irrelevant for 
the subsequent analysis we choose
\begin{equation}\label{eq:shape}
 r_\perp(z) = 
 \frac{2 C L^2}{\pi} \left(1 - \cos(\pi z / 2 L) \right) \, ,
\end{equation} 
which mimics an experiment where an end-grafted bundle is deformed by a tip-load 
at its free end \cite{trivial1}. Note that the choice of \eq{eq:shape} is not 
essential: qualitatively similar results are obtained with different bundle 
shapes also. The parameter $C$ reflects the bundle curvature and will serve as a 
measure of the amplitude of the imposed bending deformation.

In principle, crosslink reorganization may affect the local curvature and lead 
to the formation of kinks in the bundle contour~\cite{cohenPNAS2003, 
DogicPRL2006Kinks}. In the following we are primarily interested in the effect 
of an imposed deformation on the crosslink binding state, thus, bundle shape is 
assumed to be given and constant over the time-scale of interest. To bring the 
bundle from configuration $1 \to 2$ obviously requires a bending energy, $W_{\rm 
bend} \propto \KF \int_0^L ( \partial^2 r_\perp(z) / \partial z^2 )^2 
\, dz$, with $\KF$ the filament bending stiffness. However, as $r_\perp(z)$ is 
not allowed to change $W_{\rm bend}$ plays no role in what follows.

The inner structure of the bundle is an array of $i=1,\ldots,\NF$ parallel 
filaments that are spaced a distance $b$ apart (\fig{fig:illustration1}(b)). 
Each filament is a chain of $\alpha=1,\ldots,\NX$ beads, with harmonic springs 
(horizontal lines) joining nearest neighboring beads; the spring constant equals 
$\KS$, and $a$ denotes the equilibrium spring length. Beads can only slide along 
the contour, making their motions effectively one-dimensional. It therefore 
suffices to assign a single number $u_{i,\alpha}$ to each bead, denoting the 
relative displacement of that bead from its equilibrium position.  The 
possibility of performing lateral motion transverse to the bundle axis is 
thereby neglected. At sufficiently low crosslink density the entropy 
stored in these bending degrees of freedom have been shown to drive an 
unbundling transition\cite{benetatosPRE2003, kierfeld05PRL}. Here, we are 
primarily interested in highly crosslinked bundles away from the unbundling 
transition, such that the lateral degrees of freedom can be assumed to be frozen 
out.

The filaments are joined to each other by crosslinks (vertical bonds in
Fig.\ref{fig:illustration1}(b)). A pair of beads can be crosslinked provided
they are vertical nearest neighbors, i.e.~a bead $(i,\alpha)$ can only be
crosslinked to the two beads $(i \pm 1,\alpha)$ and not to any other beads. The
maximum number of crosslinks thus equals
\begin{equation}\label{eq:nmax}
 N_{\rm max} = (\NF-1) \NX,
\end{equation}
but we emphasize that not all \ahum{allowed} vertical bonds are necessarily 
crosslinked, and so the actual number of crosslinks $N$ will generally be lower.

The two dominant contributions to the elastic energy of the bundle are an axial
strain and a shear strain. The former is due to stretching of
filaments and may be written as
\begin{equation}\label{eq:hs}
 H_s = \frac{\KS}{2} \sum_{i=1}^{\NF} \sum_{\alpha=2}^{\NX} 
 \left( u_{i,\alpha-1} - u_{i,\alpha}  \right)^2,
\end{equation}
with $u_{i,\alpha}$ the relative bead displacements, and $\KS$ the spring 
constant of the horizontal bonds defined previously. Note that $\KS$ is 
related to real material properties via $\KS=EA/a$, where $E$ is the 
filament Young modulus, $A$ its cross-sectional area, and $a$ the spacing 
between successive sites along the filament backbone. 

The resistance to shear deformations is mediated by the crosslinks: with their 
two heads crosslinks connect two neighboring filaments and provide a means of 
mechanical coupling between them. While the form of the shear strain follows 
naturally from the basic definitions of continuum elasticity, it is nevertheless 
illustrative to discuss its physical basis. As a consequence of bundle 
deformation, encoded by $r_\perp(z)$ of \eq{eq:shape}, filaments have to slip 
relative to each other, bringing the crosslinking sites out of registry and 
therefore leading to crosslink deformation. The slip is given by 
$b\theta_\alpha$, where
\begin{equation}\label{eq:slip}
 \theta_\alpha = \left. \frac{d r_\perp(z)}{dz} \right|_{z=\alpha L/\NX} = 
 a C \NX \sin (\pi \alpha / 2 \NX),
\end{equation}
is the local tangent angle of the bundle at the site of the crosslink $\alpha$. 
Bringing the crosslinking sites back into registry is possible if one of the 
filaments stretches out farther than its connected partner, $u_{i+1,\alpha} = 
u_{i,\alpha} + b\theta_\alpha$, in order to compensate for the bending induced 
mismatch. The shear contribution to the elastic energy may thus be written as
\begin{equation}\label{eq:hx}
 H_\times = \frac{\KX}{2} \sum_{i=1}^{\NF-1} \sum_{\alpha=1}^{\NX} 
 n_{i,\alpha} \left( u_{i+1,\alpha} - u_{i,\alpha} + 
 b\theta_\alpha \right)^2 \, ,
\end{equation}
with $\KX$ the crosslink shear stiffness. In the above, $n_{i,\alpha}=(0,1)$ are 
the crosslink occupation variables: $n_{i,\alpha}=1$ means that between beads 
$(i,\alpha)$ and $(i+1,\alpha)$ a crosslink exists, while $n_{i,\alpha}=0$ means 
that no crosslink is present. A key ingredient of this work is the possibility 
of crosslink (un)binding: the crosslink occupation variables $n_{i,\alpha}$ are 
therefore fluctuating quantities (in contrast to previous 
studies\cite{heussingerPRL2007WLB, heussingerPRE2010} where they were quenched). 
We emphasize\cite{loveBOOK} that \eq{eq:hx} can also be derived by properly 
discretizing the shear energy of an elastic continuum with shear 
modulus~$G=\KX/a$.

The bundle model of the present study is thus defined by the Hamiltonian
\begin{equation}\label{eq:model}
 H_{\rm bundle} = H_s + H_\times \, ,
\end{equation}
i.e.~the sum of the stretch and shear contributions. Hence, there is a 
competition between filament stretching and crosslink shearing which provides 
the fundamental physical mechanism that governs the phenomena to be described. 
In the sections to come, we will study \eq{eq:model} using mostly the grand 
canonical ensemble.  That is, we fix the crosslink chemical potential $\mu$, but 
the total number of crosslinks $N = \sum_{i=1}^{\NF-1} \sum_{\alpha=1}^{\NX} 
n_{i,\alpha}$ is allowed to fluctuate.

\subsection{Units and conventions}

The key parameters in our model are the bending amplitude $C$, the crosslink 
chemical potential $\mu$, the number of filaments $\NF$, and the spring 
constants $\KS,\KX$. We also introduce the crosslink density $n=N/N_{\rm 
max}$, with $N$ the number of crosslinks between the filaments, and $N_{\rm 
max}$ the maximum number possible (see \eq{eq:nmax}). In what follows we choose 
$\beta\KS =100$, but the ratio $\KR$ will be varied (irrelevant factors of 
$\beta=1/k_BT$ are thus absorbed in the spring constants). We also expect a 
dependence on the bundle length $L=a \NX$, especially near phase transitions. 
The lattice constants $a,b$ are set to unity. We consider an end-grafted bundle, 
corresponding to the boundary condition $u_{i,1}=0$. All other displacements 
($u_{i,\alpha>1}$), as well as the bond occupation variables ($n_{i,\alpha}$), 
are fluctuating quantities.

\section{Computer Simulation Results}
\label{sec:sim}

The simulations are performed using grand canonical Monte Carlo 
\cite{frenkel.smit:2001} combined with an umbrella sampling scheme 
\cite{virnau.muller:2004} (see Appendix~A). The key output is the (normalized) 
distribution $P(N)$, defined as the probability to observe the bundle in a state 
with $N$ crosslinks. The umbrella sampling scheme ensures that $P(N)$ is 
measured over the entire range $0 \leq N \leq N_{\rm max}$, even in regions 
where $P(N)$ is very small. As a consequence, in the vicinity of a first-order 
phase transition, our results are less susceptible to hysteresis 
\cite{citeulike:4503186, berg.neuhaus:1992}.

\subsection{The case $\NF=2$}

\begin{figure}
\begin{center}
\includegraphics[width=0.7\columnwidth]{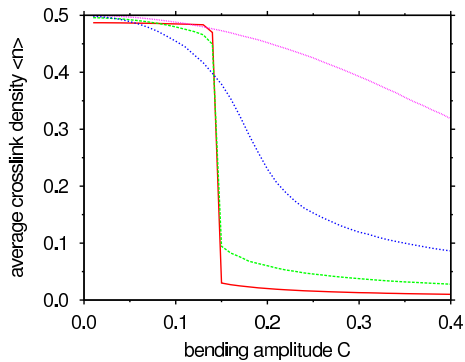}
\caption{\label{fig2} Behavior of the bundle upon bending. Plotted is the 
average crosslink density $\avg{n}$ versus bending amplitude $C$ for $\KR = 
10^{-5}, 10^{-4}, 10^{-3},10^{-2}$ (from top to bottom). For low values of 
$\KR$, $\avg{n}$ decreases smoothly with $C$; for higher values, a strong 
first-order transition is observed (data for $\NF=2, \NX=150, \mu=0$).}
\end{center} 
\end{figure}

We begin our simulations with a bundle consisting of $\NF=2$ filaments. In 
\fig{fig2}, we plot the average crosslink density $\avg{n} = \sum N P(N) / 
N_{\rm max}$, as function of the imposed bending amplitude~$C$ for several 
values of $\KR$. In all cases, $\avg{n}$ decreases with $C$, showing that the 
crosslinks unbind upon bending. The striking feature is that, for $\KR$ high 
enough, $\avg{n}$ drops extremely sharply at some special value of the bending 
amplitude. In this situation, the unbinding of crosslinks is a collective 
phenomenon, reminiscent of a first-order phase transition.

\begin{figure}
\begin{center}
\includegraphics[width=0.7\columnwidth]{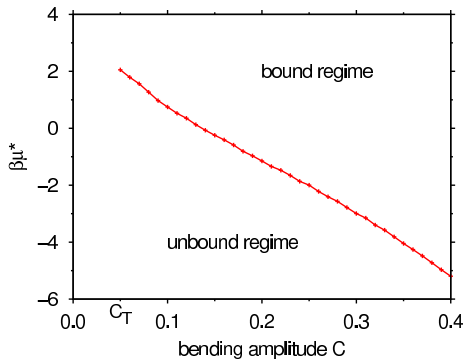}
\caption{\label{fig3} Phase diagram of a bundle consisting of $\NF=2$ filaments. 
Plotted is the chemical potential $\mu^\star$ of the first-order transition 
versus the bending amplitude (data for $\KR=0.01, \NX=300$).}
\end{center}
\end{figure}

We now specialize to $\KR=0.01$, i.e.~the largest value considered in 
\fig{fig2}, where the transition is distinctly first-order. For a given bending 
amplitude~$C$, we calculate the chemical potential $\mu^\star$ where the 
first-order transition occurs. To locate $\mu^\star$, $\mu$ is varied at 
fixed~$C$ until the \ahum{susceptibility} $\chi = (\avg{N^2} - \avg{N}^2)/N_{\rm 
max}$ reaches a maximum, i.e.~we numerically solve
\begin{equation}\label{eq:mu}
 \mu^\star(C) : \chi \to \mbox{max}.
\end{equation}
The \ahum{phase diagram} of \fig{fig3} shows $\mu^\star$ versus $C$ obtained in 
this way. This curve plays the role of a binodal: it separates the regime where 
the filaments are tightly bound by many crosslinks from the regime where they 
are only loosely coupled by few crosslinks. Note that the {\it simulated} 
binodal does not extend all the way $C \to 0$ but is \ahum{cut-off} at some 
threshold value $C=C_T$, which reflects the finite length of the bundle. Hence, 
$C_T$ does not correspond to a critical point in the usual thermodynamic 
sense\cite{trivial2}.

\begin{figure} 
\begin{center} 
\includegraphics[width=0.7\columnwidth]{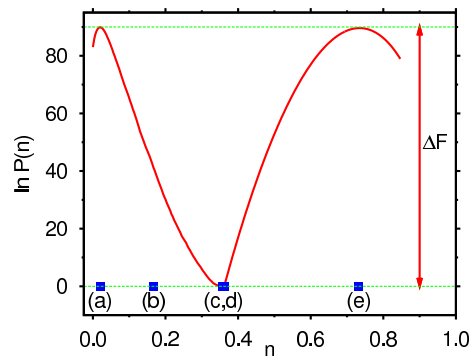} 
\caption{\label{fig:pn} $\ln P(n)$ at $\mu^\star$ of the susceptibility maximum 
(conform \eq{eq:mu}). The left (right) peak corresponds to an unbound (bound) 
bundle, while at intermediate densities the bundle is partially bound. Note the 
large free energy barrier $\Delta F$ separating the bound and unbound states 
(vertical arrow). The squares (a-e) indicate the densities $n$ at which the 
crosslink profiles of \fig{fig:prof} were measured (data for $\NF=2, \NX=300, 
\KR=0.01, C=0.195$).}
\end{center} 
\end{figure}

\begin{figure}
\begin{center}
\includegraphics[width=0.7\columnwidth]{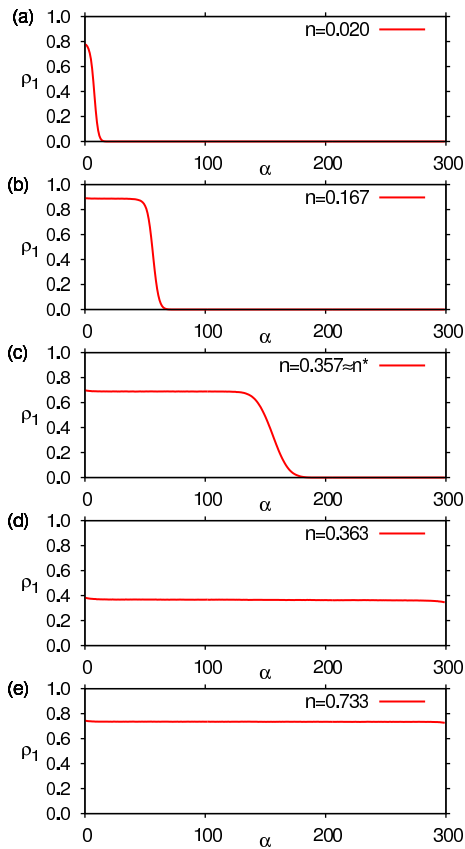}
\caption{\label{fig:prof} Crosslink density profiles $\rho_1(\alpha)$ for 
several values of the overall crosslink density~$n$. The values of $n$ in (a,e) 
coincide with the peak positions in $P(n)$ of \fig{fig:pn} and thus reflect the 
\ahum{coexisting} unbound and bound states. By increasing~$n$, a domain wall 
gradually shifts toward the bundle center (b,c) up to a density $n^\star$ where 
it \ahum{jumps} to the bundle end (d). For the parameters used here $(\NF=2, 
\NX=300, \KR=0.01, C=0.195)$ we find $n^\star \approx 0.36$, which coincides 
with the crosslink density where $P(n)$ attains its minimum.}
\end{center}
\end{figure}

Next, we investigate how the crosslinks unbind at the transition (i.e.~when 
$\mu=\mu^\star$). In \fig{fig:pn}, we show the {\it logarithm} of the 
distribution $P(n)$ measured at $\mu^\star$ (note that $\ln P(n)$ may be 
regarded as {\it minus} the free energy of the bundle). The distribution is 
distinctly bimodal, as is characteristic of a first-order transition 
\cite{citeulike:3717210}. In addition, we have checked that the barrier $\Delta 
F$ increases with the bundle length \cite{trivial3}, providing further 
confirmation that the transition is genuinely first-order 
\cite{citeulike:3908342}.

To understand how the transition from the unbound to the bound state progresses, 
we associate features of the distribution $P(n)$ to the spatial organization of 
crosslinks within the bundle. To this end, we introduce the crosslink density 
profile $\rho_i(\alpha) \equiv \avg{n_{i,\alpha}}$ measured along the bundle 
contour $\alpha=1,\ldots,\NX$ between \ahum{adjacent} filaments $i$ and $i+1$. 
Of course, for $\NF=2$, there is only one such profile: $\rho_1(\alpha)$. Some 
typical profiles are shown in \fig{fig:prof}, each obtained for a different 
value of the overall crosslink density~$n$. In~(a), we show $\rho_1(\alpha)$ for 
$n=0.020$, corresponding to the left peak in $P(n)$ where the bundle is unbound. 
We observe that $\rho_1(\alpha)>0$ only in a small region near the grafted end, 
but rapidly decays to zero thereafter. Hence, there is an interface (domain 
wall) in the system, separating a region of high crosslink density from one of 
low crosslink density. As $n$ increases, the domain wall gradually shifts toward 
the center of the bundle (b,c) up to some threshold density $n^\star$. At 
$n^\star$, the domain wall jumps discontinuously toward the bundle end, yielding 
a constant crosslink density along the entire contour (d). The value of 
$n^\star$ is given by the crosslink density where $P(n)$ attains its minimum. 
Once the domain wall has \ahum{jumped}, increasing~$n$ further no longer affects 
the shape of the profile but merely raises the plateau value~(e).

\begin{figure}
\begin{center}
\includegraphics[width=0.7\columnwidth]{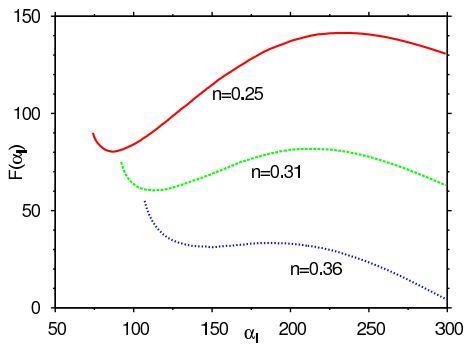}
\caption{\label{fig:ip} Bundle free energy $F(\IP)$ as function of the domain 
wall position $\IP$ for three values of the crosslink density~$n$. As $n$ 
increases, a first-order transition takes place at which the domain wall 
\ahum{jumps} from $\IP<\NX/2$ to $\IP=\NX$ (data for $\NF=2, \NX=300, \KR=0.01, 
C=0.195$).}
\end{center}
\end{figure}

The fact that the domain wall \ahum{jumps} at $n^\star$ indicates another 
first-order transition. To make this explicit, we performed a number of 
canonical simulations (i.e.~at fixed $n$), and measured the bundle free energy 
$F(\IP)$ as function of the domain wall {\it position}~$\IP$ (in simulations 
$\IP$ is set by the crosslink furthest away from the grafted end). The result is 
shown in \fig{fig:ip} for three values of the crosslink density~$n$. In all 
cases, $F(\IP)$ reveals two minima: the minimum at $\IP<\NX/2$ ($\IP=\NX$) 
corresponds to the unbound (bound) state. Note that the overall shape of 
$F(\IP)$ rather resembles a cubic polynomial in $\IP$, which is the standard 
form of the Landau free energy expansion to describe a first-order transition 
($\IP$ being the order parameter, and $n$ the temperature). For small $n$, the 
unbound state is stable (top curve). As $n$ increases, a first-order transition 
takes place above which the bound state is stable (lower curve). Precisely at 
the transition, the minima have the same free energy, here at $n \approx 0.31$ 
(center curve). Note that this somewhat underestimates $n^\star \approx 0.36$ of 
\fig{fig:prof}. One reason for the discrepancy is the use of different ensembles 
(canonical versus grand-canonical) in systems of finite size. Another reason is 
that the (un)bound states remain meta-stable over a considerable range in $n$ 
around the transition. To accurately obtain $n^\star$, one would need to perform 
an umbrella sampling simulation of the full two-dimensional probability 
distribution $P(N,\IP)$.

\subsection{The case $\NF=4$}

\begin{figure*}
\begin{center}
\includegraphics[width=2\columnwidth]{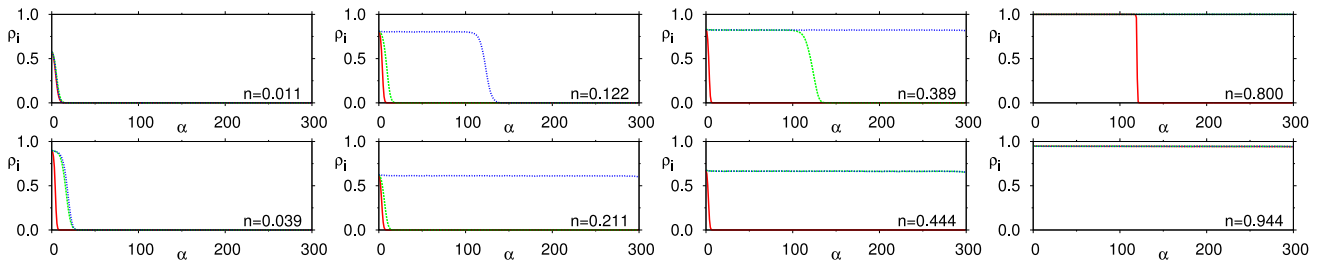}
\caption{\label{fig:qp} Simulation evidence for the existence of a sequence of 
binding/unbinding transitions in a bundle consisting of $\NF=4$ filaments. 
Plotted are the crosslink density profiles $\rho_i(\alpha)$ for several values 
of the crosslink density $n$. In each of the graphs, the solid curve corresponds 
to $\rho_2(\alpha)$ of the center filament pair; dotted and dashed curves 
represent $\rho_1(\alpha)$ and $\rho_3(\alpha)$ of the outer pairs (data for 
$\NX=300$, $\KR=0.01$, $C=0.19$).}
\end{center}
\end{figure*}

\begin{figure}
\begin{center}
\includegraphics[width=0.7\columnwidth]{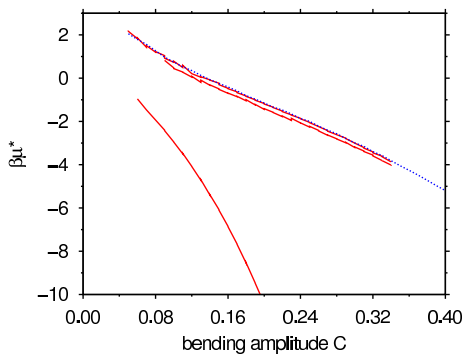}
\caption{\label{fig:bin4} Phase diagram of a bundle consisting of $\NF=4$ 
filaments (solid curves). Plotted is the chemical potential $\mu^\star_I$ of the 
$I=1,2,3$ first-order phase transitions versus the bending amplitude; the deep 
lower curve corresponds to the binding of the center filament pair (data for 
$\KR=0.01, \NX=200$). For completeness, the binodal for $\NF=2$ is also shown 
(dotted curve). As in \fig{fig3}, the binodals do not extend to $C \to 0$ due to 
finite bundle length.}
\end{center}
\end{figure}

More generally, for a bundle consisting of $\NF$ filaments, we expect a {\it 
sequence} of $\NF-1$ first-order transitions, one for each pair of adjacent 
filaments. To characterize these transitions, crosslink density profiles were 
measured for a bundle with $\NF=4$. In this case, there are $i=1,2,3$ adjacent 
filament pairs, with corresponding density profiles $\rho_i(\alpha)$. The 
profiles are depicted in \fig{fig:qp} for several values of the crosslink 
density~$n$. For small $n$, the bundle is unbound: $\rho_i(\alpha)$ is zero 
everywhere, except for a small region near the grafted end ($n=0.011$). As $n$ 
increases, the crosslinks preferentially bind the outer filament pairs 
($i=1,3$), while the center pair ($i=2$) remains unbound ($n=0.039$). Increasing 
$n$ further, the initial symmetry between outer pairs gets broken: with equal 
probability, one of the outer pairs $i=1,3$ is selected; the binding of 
crosslinks then continues for that pair only ($n=0.122$) ultimately leading to 
the first transition of the sequence ($n=0.211$). After the first transition, we 
thus have a bundle where one of the outer filament pairs is bound, while the 
remaining two pairs are unbound. Next, the other outer pair begins to bind 
($n=0.389$) leading to the second transition ($n=0.444$). We now have a bundle 
where both outer filament pairs are bound, and the \ahum{$1 \leftrightarrow 3$} 
symmetry is restored again. Not surprisingly, the third (and last) transition of 
the sequence involves the binding of the center filament pair. The mechanism is 
the same as before, featuring a domain wall ($n=0.800$) that \ahum{jumps} at the 
transition ($n=0.944$).

Note that the binding transitions of the outer filament pairs occur relatively 
close to each other (the corresponding densities are $n_1 \sim 0.2$ and $n_2 
\sim 0.4\approx 2n_1$, respectively). However, to induce the binding of the 
center pair, a significantly larger density $n_3 \sim 0.9$ is required. This 
becomes more pronounced in the $(C,\mu)$~phase diagram. To each transition 
$I=1,2,3$ in the sequence corresponds a (local) maximum in the susceptibility, 
and so the transition chemical potential $\mu^\star_I$ can be evaluated via 
\eq{eq:mu} as before. The resulting phase diagram now features three binodals, 
with those corresponding to the binding of the outer filament occurring very 
close together (\fig{fig:bin4}). Note also that the first of these \ahum{outer} 
binding transitions coincides with the binodal of the $\NF=2$ bundle.

\section{Theory}
\label{sec:the}

We now present a theoretical description that captures most of the features 
observed in the Monte Carlo simulations.

\subsection{The case $\NF=2$ without bending}

We first consider a two-filament bundle ($\NF=2$) without external deformation 
($C=0$). Assume all crosslinks to be bound for the moment, $n_{1,\alpha}=1$, 
with $\alpha=1,\ldots,\NX$. The Hamiltonian of \eq{eq:model} then becomes
\begin{equation}
 H_{\rm bundle} = \frac{\Gamma_1}{2} \Delta_1^2 + \frac{1}{2}
 \sum_{\alpha=1}^{\NX} \left[ \frac{\KS}{2} (\Delta_{\alpha+1} - 
 \Delta_\alpha)^2 + \KX \Delta_\alpha^2 \right] \,, \nonumber
\end{equation}
with $\Delta_\alpha \equiv u_{2,\alpha} - u_{1,\alpha}$ and $\Gamma_1 \equiv 
\KS/2$. Strictly speaking, we also need to include the combination 
$\Sigma_\alpha \equiv u_{2,\alpha} + u_{1,\alpha}$. However, as the latter do 
not couple to the crosslink occupation variables, we need not consider them in 
our treatment.

The idea is to iteratively integrate out the degrees of freedom 
$\Delta_\alpha$, and to monitor the resulting \ahum{flow} of the coupling 
constant~$\Gamma$. Each time one of the $\Delta_\alpha$ variables is integrated 
out, the Hamiltonian retains the above form, but with a renormalized coefficient 
$\Gamma$ given by the recursion relation
\begin{equation}\label{eq:flow.k}
 \Gamma_{i+1} = 
 \frac{\KS}{2}\frac{\KX+\Gamma_i}{\KX+\KS/2+\Gamma_i} \,,
\end{equation}
with the fixed-point
\begin{eqnarray}\label{eq:fixedpoint.k}
 \Gamma_\infty =  \frac{\KX}{2} \left(
 \sqrt{1+2\KS/\KX} -1 \right) \,.
\end{eqnarray}
At the $i$-th iteration step, the partition function $Z$ thus acquires a factor 
$(\KX+\KS/2+\Gamma_i)^{-1/2}$, such that one can write 
\begin{eqnarray}\label{eq:Zfully.bound}
 Z = e^{-\beta\mu \NX} \prod_{i=1}^{\NX} 
 \left( \frac{\D \KX + \Gamma_i}{\D \Gamma_{i+1}}\right)^{-1/2},
\end{eqnarray}
where we have used \eq{eq:flow.k} and dropped an overall factor 
$(\KS/2)^{-\NX/2}$.

To see how the above generalizes to the case of open crosslinks let us assume 
that the crosslinks from sites $\alpha=j,\ldots,j+l_j-1$ are open. We will call 
this a \ahum{bubble} of length $l_j$ in the following. The associated variables 
$\Delta_\alpha$ can immediately be integrated over, with the effect of 
generating a term
\[
 (\KS^{\rm eff}/2)(\Delta_{j+l_j}-\Delta_{j-1})^2, \quad
 \KS^{\rm eff} \equiv \KS/(l_j+1),
\] 
in the renormalized Hamiltonian. Thus, to account for bubbles, we have to 
substitute the stretching stiffness $\KS$ with $\KS^{\rm eff}$ in 
Eqs.~(\ref{eq:flow.k}) and (\ref{eq:Zfully.bound}), as well as to reinterpret 
$\NX$ as the number of {\it bound} sites: $N_{\rm bound} \equiv N$. The 
resulting expression is an exact, albeit intractable, representation of the 
partition function. 

To make progress, we use a mean-field (MF) approximation, where we assume the 
crosslinks to be homogeneously distributed along the bundle (the interface will 
be accounted for later). The actual bubble length thus gets replaced by its 
average value $1/(l_j+1) \to N/\NX \equiv x$, where $x$ denotes the fraction of 
bound crosslinks. Furthermore, we replace the renormalized coupling constant by 
its fixed-point value, $\Gamma_i \to \Gamma_\infty$, which in our 
MF~approximation can be written as
\[
 \Gamma_\infty(x) = \frac{\KX}{2} \left(
 \sqrt{1 + 2x \KS/\KX} - 1 \right). 
\]
Within these approximations, the partition function of \eq{eq:Zfully.bound} 
evaluates to 
\begin{eqnarray}\label{eq:Z.mf.final}
 Z_{\rm MF} = \sum_{N=0}^{\NX} p_N \, e^{-\beta\mu N}
 \left(1 + \frac{\KX}{\Gamma_\infty (N/\NX)} \right)^{-N/2} \,,
\end{eqnarray}
which can easily be evaluated numerically. The term $p_N=\binom{\NX}{N}$
represents the usual \ahum{entropy of mixing} and counts the number of crosslink
configurations compatible with a given $N = x \NX$. We will show in Appendix~B
how the averaged crosslink occupation $\avg{n}$ compares to our simulation
results. We will furthermore show how one can improve the theory by explicitly
incorporating bubbles using the \ahum{necklace model} of Fisher
~\cite{fisherJStatPhys1984Necklace,husePRB1984Necklace}.

\subsection{The case $\NF=2$ with bending and interface}

Next, we incorporate a finite bending amplitude $\theta_\alpha$ into the theory. 
In addition, as the simulations indicate the possibility of an interface between 
a region of high and low crosslink density, we must appropriately generalize the 
above MF~approach to the crosslink occupation variables $n_{1,\alpha}$. 
To this end, we assume the crosslinks to be homogeneously distributed in the 
region of high density only. A sharp interface separates this region from one 
without crosslinks
\begin{eqnarray}\label{eq:Ansatz.n}
 n_{1,\alpha} = \begin{cases}
 xN_\times/\IP   &  \alpha<\IP, \\
 0 &  \alpha>\IP,
\end{cases}
\end{eqnarray}
where $\IP \in [0,N_\times]$ is the (unknown) axial position of the interface. 
Below we will also use the {\it normalized} interface position
\begin{equation}\label{eq:yip}
y=\IP/N_\times \in [0,1] \,.
\end{equation}
As before, $x$ denotes the fraction of bound crosslinks, which can be expressed
in terms of the occupation variables as $x=\sum_\alpha n_{1,\alpha}/\NX$.

We now make an \ahum{Ansatz} for the displacement degrees of freedom 
$u_{i,\alpha}$. We assume that the quadratic terms in \eq{eq:hx} are small 
whenever there are crosslinks that bind the two filaments together. That is, 
provided $n_{1,\alpha}=1$, the corresponding displacement $\Delta_\alpha \equiv 
u_{2,\alpha} - u_{1,\alpha} \propto \theta_\alpha$. In the region of low 
crosslink density we can assume that $\Delta_\alpha=\rm const$. We thus obtain
\begin{eqnarray}\label{eq:Ansatz.Delta}
 \Delta_\alpha = \begin{cases}
 u_0 \sin(\pi \alpha/2 \NX) & \alpha<\IP, \\
 u_0 \sin(\pi \IP/2 \NX) & \alpha>\IP,
\end{cases}
\end{eqnarray}
by requiring continuity at $\alpha=\IP$, and where also \eq{eq:slip} was 
used. Introducing these expressions into the Hamiltonian of \eq{eq:model}, and 
minimizing with respect to $u_0$, we obtain the following \ahum{saddle-point} 
contribution to the effective free energy
\begin{eqnarray}\label{eq:Fmodel}
 F_{\rm sp}(x,y) = \NX Af_c(y)\left[1 + \D\frac{x_0}{x}\cdot\frac{yf_c(y)}{f_s(y)}
 \right]^{-1} \,,
\end{eqnarray}
with functions
\begin{equation}
\begin{split}
 f_c(y)= \frac{2}{L}\int_0^{yL} \cos^2(\pi s/2)L \, ds\,, \\
 f_s(y)= \frac{2}{L}\int_0^{yL} \sin^2(\pi s/2L) \, ds \,,
\end{split}
\end{equation}
and bundle length $L=a\NX$. The relevant parameters are $A \simeq \KS 
b^2(Ca)^2$, which encodes the dependence on bending amplitude $A\propto C^2$, 
and $x_0\simeq (k_s/k_\times)/N_\times^2$ representing the effects of the 
crosslink stiffness $x_0\propto 1/k_\times$. The point to note is that 
$F_{\rm sp}(x,y)$ still depends on the crosslink occupation variable~$x$, as 
well as on the location of the interface~$\IP$ (via the parameter $y$ of 
\eq{eq:yip}).

\begin{figure}
\begin{center}
\includegraphics[clip=,width=0.9\columnwidth]{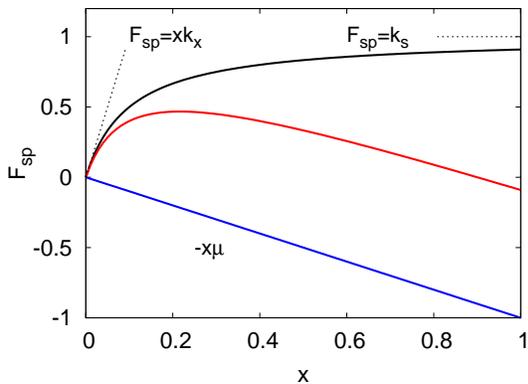}
\caption{\label{Fsp.simple.picture} Illustration of the mechanism that induces a 
discontinuous reduction of the crosslink occupation as function of bending 
amplitude and/or chemical potential. The top curve shows $F_{\rm sp}(x)$ of 
\eq{eq:Fmodel.z1}, with the dashed lines indicating the limiting behaviors for 
$x \to 0$ and $x \to 1$. The binding enthalpy $-x\mu$ (lower line), which is to 
be added to $F_{\rm sp}(x)$, \ahum{tilts} the free energy landscape, leading to 
a free energy (center curve) featuring a coexistence between two states (at high 
and low crosslink occupation~$x$, respectively). As the control parameters 
$A,\mu$ are varied, a discontinuous transition from one state to the other is 
thus observed.}
\end{center}
\end{figure}

In a previous report \cite{PhysRevE.83.050902} we 
presented the special case $y=1$, i.e.~without an interface being present. In 
this limit one obtains for the free energy
\begin{eqnarray}\label{eq:Fmodel.z1}
 F_{\rm sp}(x,y=1) \equiv F_{\rm sp}(x) = N\frac{A}{1+x_0/x} \,.
\end{eqnarray}
This simple form, which is illustrated in \fig{Fsp.simple.picture}, conveys an 
intuitive picture of how a finite bending amplitude may lead to a discontinuous 
reduction of crosslink occupation $x$. When $x \ll x_0$ the free energy 
essentially grows linearly, $F_{\rm sp}(x) \sim x \KX$, with the energy scale 
set by the crosslink stiffness $\KX$. This indicates that each crosslink 
contributes a certain amount of deformation energy, while the filaments remain 
nearly undeformed. The few crosslinks present are just not strong enough to 
force the filaments into a deformed state. This situation changes when $x \gg 
x_0$, where the free energy saturates at a value set by the filament stretching 
stiffness, $F_{\rm sp}(x) \sim \KS$. Now there are enough crosslinks to stretch 
out the filaments, at the same time relieving their own deformation. Together 
with the binding enthalpy, $E=-x \mu$, which leads to the usual tilting of the 
free energy landscape, we obtain a total free energy that has two coexisting 
states, at high and low crosslink occupation. As the bending amplitude $A$ or 
the chemical potential $\mu$ is varied, it is therefore possible to observe a 
discontinuous transition from one state to the other.

\begin{figure}
\begin{center}
\includegraphics[clip=,width=0.95\columnwidth]{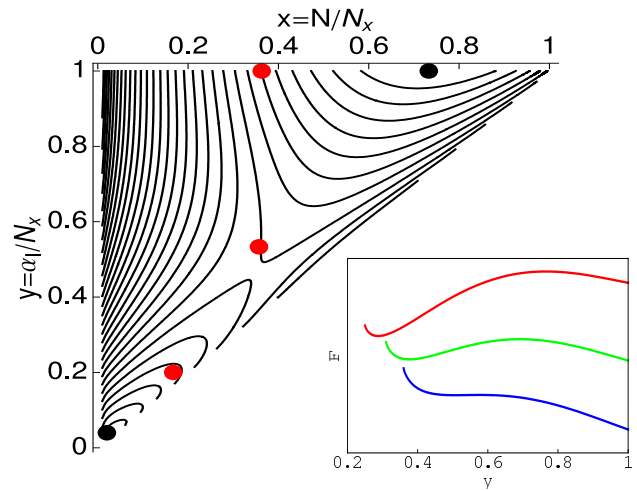}

\caption{\label{Fsp.contour} Contour plot of the total free energy $F(x,y)$ 
given by \eq{eq:Fall}. Note that only the region $x<y$ is physically relevant as 
the total number of crosslinks $N$ is always less than~$\alpha_I$. The dots 
correspond to the state points (a-e) in \fig{fig:prof} where the crosslink 
profiles were obtained in simulations. The inset shows the free energy $F$ as 
function of the normalized interface position $y$ for {\it fixed} 
$x=0.25,0.31,0.36$ (from top to bottom), which compares well to the simulation 
result of \fig{fig:ip}.}

\end{center}
\end{figure}

\begin{figure}
\begin{center}
\includegraphics[clip=,width=0.7\columnwidth]{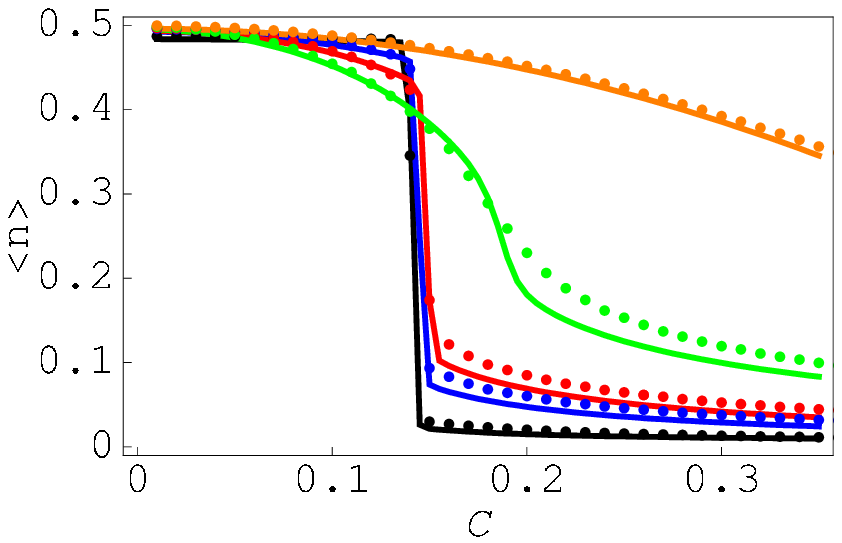}
\includegraphics[clip=,width=0.7\columnwidth]{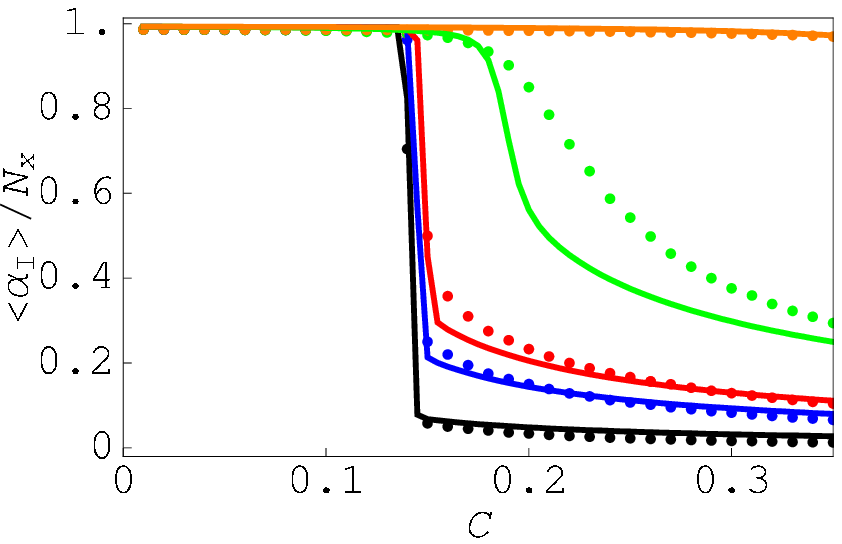}
\caption{\label{avg.n.nf2} (top) Average crosslink occupation $\avg{n}$ and
  (bottom) average interface location $\avg{\IP}$ as function of bundle
  curvature $C$ using $\NX=150$, and for different crosslink stiffness $\KX/\KS
  =10^{-5},10^{-4},4\cdot10^{-4},10^{-3},10^{-2}$ (from top to bottom). In both graphs, the lines are
  obtained without fit parameters from the theoretical model, \eq{eq:Fall},
  while the symbols indicate simulation data.}
\end{center}
\end{figure}

The existence of an interface does not change fundamentally this picture but 
adds a second reaction coordinate that the bundle can utilize in order to 
minimize its free energy during the unbinding process. Combining 
Eqs.~(\ref{eq:Z.mf.final}) and (\ref{eq:Fmodel}) we obtain $Z = 
\sum_{\IP=0}^{\NX} \sum_{N=0}^{\IP} e^{-\beta F(N,\IP)}$ with
\begin{eqnarray}\label{eq:Fall}
 e^{-\beta F(N,\IP)} &=& \\
 \left( {\IP \atop N} \right) &\times&
 \left( 1 + \frac{\KX}{\Gamma_\infty (N/\IP)} \right)^{-N/2}
 \times e^{-\beta \left(F_{\rm sp} + \mu N \right)}, \nonumber
\end{eqnarray}
which defines the total free energy $F(N,\IP)$. The latter is illustrated in
\fig{Fsp.contour} using parameters that correspond to the discontinuous
transition of \fig{fig:prof}. The figure strikingly shows the coexisting states
at high and low crosslink density, and the formation of an interface upon
passing over the intermediate saddle-point. The theory also reproduces the free
energy barrier along lines of constant~$N$ (inset), in line with the canonical
simulations of \fig{fig:ip}. The transition pathway followed in these
simulations is indicated by the light (red) points. After passing the
saddle-point, at $x\approx0.35$, the interface \ahum{jumps} from the center of
the bundle to the distant end. It is interesting to compare this
\ahum{canonical} pathway with the general shape of the basin of attraction into
the bound state. This seems to favor a pathway closer to the diagonal, with an
associated continuous motion of the interface.

From \eq{eq:Fall} it is straightforward to calculate the average crosslink 
density $\avg{n}$, as well as the average location of the interface $\avg{\IP}$. 
Both are compared to simulation results in \fig{avg.n.nf2}, and the agreement is 
remarkably good: as the bending amplitude increases, there is a discontinuous 
jump in both the crosslink density, as well as in the interface position.

\subsection{The case $\NF>2$}

Let us now turn to bundles with more than two filaments. The simulations have 
indicated a sequence of $\NF-1$ transitions, one transition for each adjacent 
filament pair (\fig{fig:qp}).  Upon increasing the bundle deformation $C$, the 
crosslinks in the central filament pair unbind first. This unbinding transition 
leaves two smaller weakly coupled sub-bundles. The successive transitions then 
happen in the centers of these sub-bundles up until all filament pairs are 
unbound.

As with each filament pair the number of variational parameters increases, a 
full theoretical treatment becomes intractable. We therefore choose a 
semi-analytic treatment that explicitly accounts for crosslinks in the 
respective central layer only. The stretching degrees of freedom that do not 
belong to this layer are integrated out by assuming\cite{heussingerPRE2010}
\begin{eqnarray}\label{eq:uia}
  u_{i+1,\alpha}-  u_{i,\alpha} = \Delta_\alpha\,,
\end{eqnarray}
independent of the layer index~$i$. This indicates that filament stretching 
increases approximately linearly with the filament index, i.e.~with the distance 
from the center of the bundle (filaments farther out from the center 
\ahum{inherit} the stretching from all the filaments on the inside). Such a 
linear dependence constitutes a central assumption in classical continuum 
theories, such as Euler-Bernoulli or Timoshenko beam theories~\cite{timoshenko}. 
For sufficiently stiff crosslinks, and without considering the possibility of an 
interface ($y=1$), this model can be mapped onto a two-filament bundle, conform 
\eq{eq:Fmodel.z1}, with effective $\NF$-dependent parameters, $A(\NF)\sim \NF^3$ 
and $x_0(\NF)\sim \NF$. We then use these $\NF$-dependent parameters in the full 
free energy of \eq{eq:Fall} to calculate the average crosslink density for the 
given layer.

\begin{figure}
\begin{center}
\includegraphics[clip=,width=0.7\columnwidth]{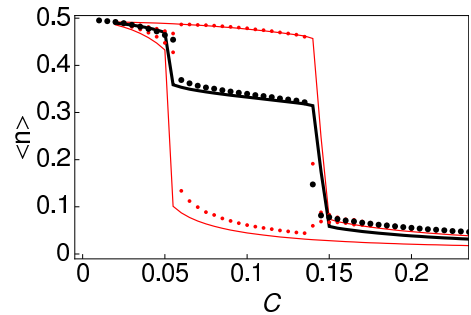}

\caption{\label{avg.n.nf2.4.6} The analogue of \fig{fig2} but for $\NF=4$. The 
key difference is that we now observe a sequence of unbinding transitions. 
Plotted is the average crosslink density $\avg{n}$ as function of the bending 
amplitude~$C$ for $\KR=10^{-3}, \, \mu=0, \, \NX=150$ (curves show results 
obtained using the \ahum{Ansatz} \eq{eq:uia}; symbols represent simulation 
data). The lower (upper) curve shows the crosslink density in the central 
(outer) layer. The middle curve reflects the crosslink density of the entire 
bundle, which was obtained by adding the contributions from the individual 
layers.}

\end{center}
\end{figure}

\fig{avg.n.nf2.4.6} compares the result of this calculation (curves) with 
simulation data (symbols) for the case $\NF=4$. The middle curve shows the 
average crosslink density $\avg{n}$ of the entire bundle versus bending 
amplitude $C$: the agreement with the simulations is quite remarkable. The lower 
curve shows the average crosslink density in the central layer, which unbinds at 
$C \approx 0.05$. Here, the theory slightly underestimates the simulation 
results, but the location of the transition is accurately reproduced. The upper 
curve shows $\avg{n}$ of the outer layer, which unbinds at a much larger bending 
amplitude $C \approx 0.15$. This curve is nearly equivalent to that of a bundle 
with $\NF=2$ filaments, in agreement with the binodal of \fig{fig:bin4}.

\section{Discussion}
\label{sec:final}

We have studied the response of a reversibly crosslinked filament bundle to an 
imposed bundle deformation. The central quantity was the average crosslink 
density $\avg{n}$ and its dependence on the imposed curvature $C$ of the bundle 
backbone. As compared to simple Langmuir adsorption, 
$\avg{n}=1/(1+e^{\beta\mu})$ at chemical potential $\mu$, one expects a 
decreasing crosslink occupation with increasing bundle deformation. The reason 
is that bundle deformation leads to a mismatch between the crosslink binding 
sites, and therefore to an elastic energy cost for binding. We found this basic 
mechanism to indeed be true, but the detailed phenomenology of crosslink 
unbinding turns out to be surprisingly rich and goes beyond a simple shift of 
the equilibrium state as would be characterized, for example, by an effective 
chemical potential.

Our main result is the possibility of a cooperative and discontinuous reduction 
of crosslink occupation as the bundle deformation is increased. The reason for 
this behavior is the competition between the energy scales of crosslink shear 
$\KX$, and filament stretch $\KS$, which is particularly efficient when the 
crosslinks are stiff. An unbinding event will then affect the force balance in 
the filaments, with the potential of influencing the bundle state far away from 
the unbinding site. On the other hand, when $\KX$ is small and the crosslinks 
soft, crosslinks unbind one after the other leading to a smooth decrease of the 
average crosslink occupation.

We have characterized in detail the discontinuous unbinding transition and 
identified the existence of an interface; the latter separates a region of high 
crosslink density from a region essentially free of crosslinks. The formation of 
an interface is a collective process in which crosslinks have to reorganize 
within the bundle and find new binding sites. A similar effect has been 
discussed in the context of the twisting of helical filament bundles, where 
crosslinks organize into certain \ahum{binding zones}~\cite{heussingerJCP2011}. 
As one increases the bundle twist away from its preferred value, these binding 
zones become shorter and shorter, thus necessitating collective reorganizations 
of many crosslinks simultaneously. As evidenced in \fig{fig:ip}, such a process 
implies the crossing of a free energy barrier. The associated time-scale of 
escape over the barrier may be much longer compared to single crosslink 
(un)binding events.

Such long time-scales may indeed be present in a recent experiment with F-actin 
bundles crosslinked by $\alpha$-actinin\cite{strehleEBPJ2011}. In the experiment 
a bundle was brought into a deformed configuration, where it was kept for either 
ten or $1000$ seconds. After this waiting time the bundle was released and its 
relaxation was monitored. For the shorter waiting time the bundle showed the 
expected exponential relaxation into the straight ground state. For the longer 
waiting time, however, the bundle did not relax back, but remained in a state 
with a considerable residual bending deformation. Apparently, upon bundle 
deformation, new binding sites become available that stabilize the bent shape by 
allowing the crosslinks to rebind in more favorable states that avoid crosslink 
straining. After releasing the bundle these crosslinks act to stabilize the bent 
contour, thus leading to a plastically deformed bundle, where the ground-state 
is no longer straight\cite{trivial5}. In line with our interpretation in terms 
of a free energy barrier, the apparent time-scale (the waiting time) necessary 
to observe bundle plasticity was considerably longer than the time required for 
(un)binding of the individual $\alpha$-actinin linkers, which is on the order of 
seconds\cite{WachsstockBPJ1993}.

Strictly speaking, our model does not allow for plastic deformation as the 
crosslinks are assumed to always bind to the same, initial binding sites: a 
crosslink at site $\alpha$ only binds to the site $\alpha$ on the next filament. 
We explicitly exclude the binding of crosslinks between non-neighboring sites, 
e.g.~between $\alpha$ and $\alpha \pm 1$. If the bending-induced mismatch, 
$b\theta_\alpha$, between the original sites at $\alpha$ is large, those 
\ahum{new} sites may actually be more favorable in terms of crosslink energy.  
Binding to new sites may then help to \ahum{freeze-in} the applied bending 
deformation leading to a plastically deformed contour. Such a model would pose a 
challenging problem for a theoretical analysis and has to be left for future 
work. Instead, we propose a simple mapping that allows us to incorporate bundle 
plasticity into the \ahum{elastic} bundle model presented in this work.

To this end, we assume the new binding sites to be optimal, in the sense that 
for the given imposed bundle contour no elastic energy cost is associated with 
the rebinding of a crosslink. Reshuffling a crosslink from its original position 
to a new site may then be conceived of as being an unbinding event at zero 
chemical potential. For simplicity, we further assume that rebinding is fast 
enough, such that all crosslinks are either bound to original sites or to new 
sites. In this case, the number of crosslinks bound to new sites, $M$, can be 
inferred without further calculation from the results presented in this work, 
$\avg{M}=N_{\rm max} - \avg{N}_{\mu=0}$. If we now release the bundle from its 
deformed state, filament elasticity will try to relax the bundle back to its 
original unbent state. The population of newly bound crosslinks, however, acts 
against this relaxation and stabilizes the bent contour. Within our assumption 
the additional contribution to the crosslink shear energy is 
\begin{eqnarray}
  H^{\rm new}_\times = \frac{\KX M}{2\NX}\sum_{\alpha=1}^{N_\times}
  \left( \Delta_\alpha + b( \theta_\alpha-\theta_\alpha^0)\right)^2 \,,
\end{eqnarray}
where we assumed $\NF=2$ for simplicity. Here, $\theta_\alpha^0$ is the tangent 
in the {\it reference} configuration at position $\alpha$ that was imposed 
during the waiting time, while $\theta_\alpha$ corresponds to the tangent 
acquired during the relaxation process.

As can be seen, the magnitude of the stabilizing effect depends on $M$, which 
depends on the waiting time. If the crosslink is stiff enough, such that a free 
energy barrier is present, fast thermalization is prevented. In this case $M 
\approx 0$ on short time-scales and the bundle will behave elastically. On 
time-scales long compared to the escape time over the barrier, the number of 
crosslinks bound to new sites reaches its equilibrium value, $M=\avg{M}$, and 
the bundle is maximally plastic. In the experiment of \olcite{strehleEBPJ2011} 
only two waiting times were accessible. It would be interesting to 
systematically change the experimental time-scale. One possibility could be to 
introduce a rate, at which bundle deformation is increased. In the context of 
protein unfolding, similar experiments\cite{dudkoPNAS2008} have proved extremely 
useful to extract information on the underlying free energy landscape.

\begin{acknowledgments}

This work is financially supported by the Emmy Noether program (VI~483/1-1) of 
the {\it Deutsche Forschungsgemeinschaft}. 

\end{acknowledgments}

\appendix

\section{Monte Carlo method}

The bundle is represented by a two-dimensional $(i=1,\ldots,\NF) \times 
(\alpha=1,\ldots,\NX)$ lattice. To each lattice site $(i,\alpha)$ a real number 
$u_{i,\alpha}$ is attached, denoting the local axial displacement. In addition, 
occupation variables $n_{i,\alpha}=(0,1)$ are attached to \ahum{vertical} 
nearest neighboring pairs $(i,\alpha)$ and $(i+1,\alpha)$. We simulate in the 
grand canonical ensemble using a biased Hamiltonian
\begin{equation}\label{eq:bias}
 H_{\rm bias} = H_{\rm bundle} + W(N),
\end{equation}
with $H_{\rm bundle}$ the \ahum{unbiased} Hamiltonian of \eq{eq:model}, and 
$W(N)$ a weight function defined on the total number of crosslinks $N$. The 
purpose of $W(N)$ is to sample all microstates $0 \leq N \leq N_{\rm max}$ with 
equal probability. To construct $W(N)$, which is {\it a priori} unknown, we use 
successive umbrella sampling \cite{virnau.muller:2004}. Once known, the 
sought-for probability distribution in the number of crosslinks follows as $P(N) 
\propto \exp(-\beta W(N))$. As Monte Carlo moves we use single bead 
displacements and crosslink binding/unbinding moves, each attempted with equal 
probability. In a displacement move, a lattice site $(i,\alpha)$ is selected 
randomly, and the current displacement $u_{i,\alpha}$ of that site is replaced 
by $u_{i,\alpha}' = u_{i,\alpha} + \delta$, with $-0.1 < \delta < 0.1$ drawn 
uniformly randomly. The new displacement $u_{i,\alpha}'$ is accepted with 
probability $P_{\rm disp} = \min \left[1, e^{-\beta \Delta H} \right]$, where 
$\Delta H$ is the energy difference between initial and final state (since 
displacements do not change $N$, both \eq{eq:model} and \eq{eq:bias} can be used 
to compute the energy difference). During a crosslink move, a vertical bond is 
selected randomly, and the corresponding occupation variable $n_{i,\alpha}$ is 
\ahum{flipped} ($n_{i,\alpha}=0$ gets replaced by 1, and vice versa). The new 
state is accepted with probability $P_{\rm xlink} = \min \left[1, e^{-\beta 
\Delta H_{\rm bias} + \beta \mu \Delta N} \right]$, with $\mu$ the crosslink 
chemical potential, $\Delta N = \pm 1$ the change in the number of crosslinks, 
and $\Delta H_{\rm bias}$ the energy difference which must now be calculated 
using the {\it biased} Hamiltonian of \eq{eq:bias}.

\section{Comparison to simple mean-field \\ theory and the necklace model}

The simple mean-field approximation developed in \eq{eq:Z.mf.final} is capable 
of accurately describing the thermodynamic properties of the bundle without 
external deformation, i.e.~$C=0$. One can slightly improve on this result by 
explicitly incorporating the bubbles along the lines of the classic 
\ahum{necklace model}~\cite{fisherJStatPhys1984Necklace,husePRB1984Necklace}.

\begin{figure}
\begin{center}
\includegraphics[clip=,width=0.9\columnwidth]{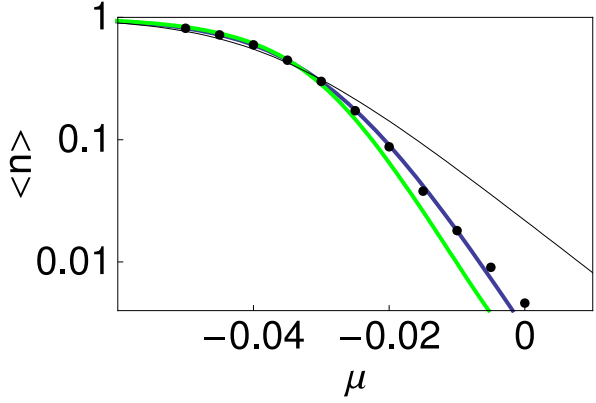}

\caption{\label{curv.zero.comparison} Average crosslink density $\avg{n}$ of an
  undeformed bundle ($C=0$) consisting of $\NF=2$ filaments as function of the
  chemical potential $\mu$ using $\KR=10^{-3}, \NX=100$. Note the logarithmic
  vertical scale! The thin curve is the solution of the one-crosslink model, as
  presented by Eq.~(5) of \olcite{PhysRevE.83.050902}, while the thick curves
  were calculated using \eq{eq:Z.mf.final} (lower curve) and the necklace model
  (middle curve).  The latter indeed yields better agreement with the simulation
  results (dots).}

\end{center}
\end{figure}

The partition function for a bubble of length $M$ is obtained from 
Eqs.~(\ref{eq:flow.k}) and (\ref{eq:Zfully.bound}) by setting $k_\times =\mu=0$
\begin{eqnarray}\label{eq:Z.bubble}
Z_0(M) = \prod_{i=1}^{M}(k_s/2+\Gamma_i)^{-1/2} = (1+\frac{M}{M_c})^{-1/2}\,,
\end{eqnarray}
where we defined a characteristic bubble size $M_c=k_s/2\Gamma_0$. The value of 
the coupling constant at the beginning of the bubble, $\Gamma_0$, can be taken 
equal to the fixed-point value in the neighboring bound segment, $\Gamma_0 = 
\Gamma_\infty(k_\times )$. The partition function of a bound segment of length 
$N$ is
\begin{eqnarray}
 Z_1(N) = \left[ e^{-\beta\mu} \left(
 1+\frac{k_\times}{k_\infty} \right)^{-1/2}\right]^N \,.
\end{eqnarray}
The full partition function can be obtained from the generating functions 
$\Phi_\alpha(z) = \sum_{N=1}^\infty z^N Z_\alpha(N)$ and the solution to the 
equation $\Phi_0(z)\Phi_1(z)=1$. From this the average crosslink density 
$\avg{n}$ is obtained in the usual way by differentiating with respect to $\mu$. 
The quality of the different approximations is compared in 
\fig{curv.zero.comparison}.

\begin{thebibliography}{10}
\bibitem{trivial1}
Strictly speaking, in \eq{eq:shape} with $C \neq 0$, one has $0 \leq z \leq
  L'$, with $L'<L$, but for small deformations $C$ the difference $L-L'$ is
  negligible.
\bibitem{trivial2}
For small $C$, \eq{eq:mu} develops a second solution at $\mu=\mu_T$, where
  $\mu_T$ is close to zero. The solution at $\mu_T$ is trivial in the sense
  that it corresponds to the \ahum{ideal} crosslink distribution $P(N) \propto
  e^{\beta \mu N}$. The threshold $C_T$ can be defined numerically as the
  bending amplitude $C$ where $\chi(\mu^\star) = \chi(\mu_T)$. As $\NX$ is
  increased, one finds that $C_T$ is well described by an empirical fit of the
  form $C_T \propto 1/\NX^{0.5}$. Hence, in the thermodynamic limit $\NX \to
  \infty$, the binodal extends all the way to $C \to 0$.
\bibitem{trivial3}
By varying the bundle length, a linear increase $\Delta F = \gamma \NX$ is
  found; the prefactor depends on the bending amplitude and is well captured by
  an empirical fit $\gamma \approx 10.3 \times C^{2.17}$.
\bibitem{trivial5}
On longer time-scales the ground-state will relax back to the straight state as
  crosslinks will again find their original binding sites. This reversibility
  has been demonstrated in the experiment.
\end{thebibliography}
\end{document}